\documentclass[aps,prl,amsmath,twocolumn,amssymb,floatfixng,showpacs,
superscriptaddress,footinbib]{revtex4}
\pdfoutput=1
\usepackage[dvips]{graphics}
\usepackage{bm}
\usepackage{float}
\usepackage{epsfig}
\usepackage{enumerate}
\usepackage{amsmath}
\usepackage{color}
\usepackage{graphicx}
\usepackage[colorlinks=true,linktoc=page,linkcolor=red,citecolor=blue]{hyperref}

\newcommand\bea{\begin{eqnarray}}
\newcommand\eea{\end{eqnarray}}
\newcommand\beq{\begin{equation}}  
\newcommand\eeq{\end{equation}}

\begin{document}

\title{\titlename}
\date{\today}
\title{Hexagonal warping induced exceptional physics in multi-Weyl semimetals}
	\author{Debashree Chowdhury}
	\thanks{These two authors contributed equally}
	\email{debashreephys@gmail.com}
		\affiliation{Centre of Nanotechnology, Indian Institute of Technology Roorkee, Roorkee, Uttarakhand-247667}
\author{Ayan Banerjee}
\thanks{These two authors contributed equally}
\email{ayanbanerjee@iisc.ac.in}
\affiliation{Solid State and Structural Chemistry Unit, Indian Institute of Science, Bangalore 560012, India}
\author{Awadhesh Narayan}
\email{awadhesh@iisc.ac.in}
\affiliation{Solid State and Structural Chemistry Unit, Indian Institute of Science, Bangalore 560012, India}
\date{\today}

\begin{abstract}
Hexagonal warping (HW) in three-dimensional topological insulators is, by now, well-known. We show that non-Hermitian (NH) loss/gain can generate an exceptional HW effect in double Weyl-semimetals (DWSM). This unique feature of DWSMs has distinctive effects on Fermi surface topology. Importantly, in the presence of such a $k^3$ spin orbit coupling mimicking term, the symmetry associated with the DWSMs is changed, leading to four exceptional points, among which two are degenerate. Introducing a driving field removes this degeneracy. The combined action of the NH warping and driving parameters leads to notable effects, including merging and tuning of exceptional points. We analyze the topological nature of the generated exceptional contours by evaluating several topological invariants, such as winding number, vorticity, and NH Berry curvature. We hope that our theoretical results would initiate possible experiments exploring NH HW effects.
\end{abstract}
\maketitle

\section{Introduction} The role of topology in condensed matter systems came to the limelight post the discovery of topological insulators (TIs) \cite{Hasan}. The presence of unique edge states and unusual Fermi surface topology, makes these systems intriguing. An important addition to the properties of the TIs is the occurrence of hexagonally warped surface states \cite{Fu,Basak} due to the presence of cubic Dresselhaus spin-orbit coupling term. In the Hermitian case, hexagonal warping (HW) is an unique property of the surface states of topological insulators, specifically for the bismuth family, where unlike the usual circular Fermi surface, the Fermi surface becomes deformed. This deformation of the Fermi surface can be well explained in terms of crystal symmetries of the surface. In Bi$_2$Se$_3$ and Bi$_2$Te$_3$ class of materials, the full rotational symmetry is absent, but due to the presence of a lower three-fold rotation symmetry, one can observe snowflake like deformed Fermi surface, whose effect can be incorporated into the Hamiltonian by adding a $k^{3}$ spin orbit coupling term to the Dirac Hamiltonian. However, three-dimensional systems such as Weyl semimetals do not show any HW effect. Besides, in three dimensions, analyzing various aspects of Dirac and Weyl semimetals (WSM) \cite{Vafek,Kane,Burkov,Armitage} are equally fascinating for the quantum matter community. WSMs are materials where the dispersion is linear and the valence and conduction bands meet at a single point, i.e., the Weyl point. The Weyl points usually behave as a source or sink of the Berry curvature with monopole charge $\pm 1$. Unlike the usual case, the multi- or $n$-WSMs do exist in nature \cite{Gupta,MW1,TW,Fang,Xu1,Huang,Zhang}, where the monopole charge can take values $n=2,3,4.$ Apart from having non-linear dispersions, these $n$-WSM preserve $C_{n}$ rotational symmetry. Analogous to TIs, WSMs also exhibit topological surface states, coined as the Fermi-arcs, which exhibit numerous interesting properties \cite{Cerjan1}. 
\begin{figure}	
	\includegraphics[width=0.95\linewidth]{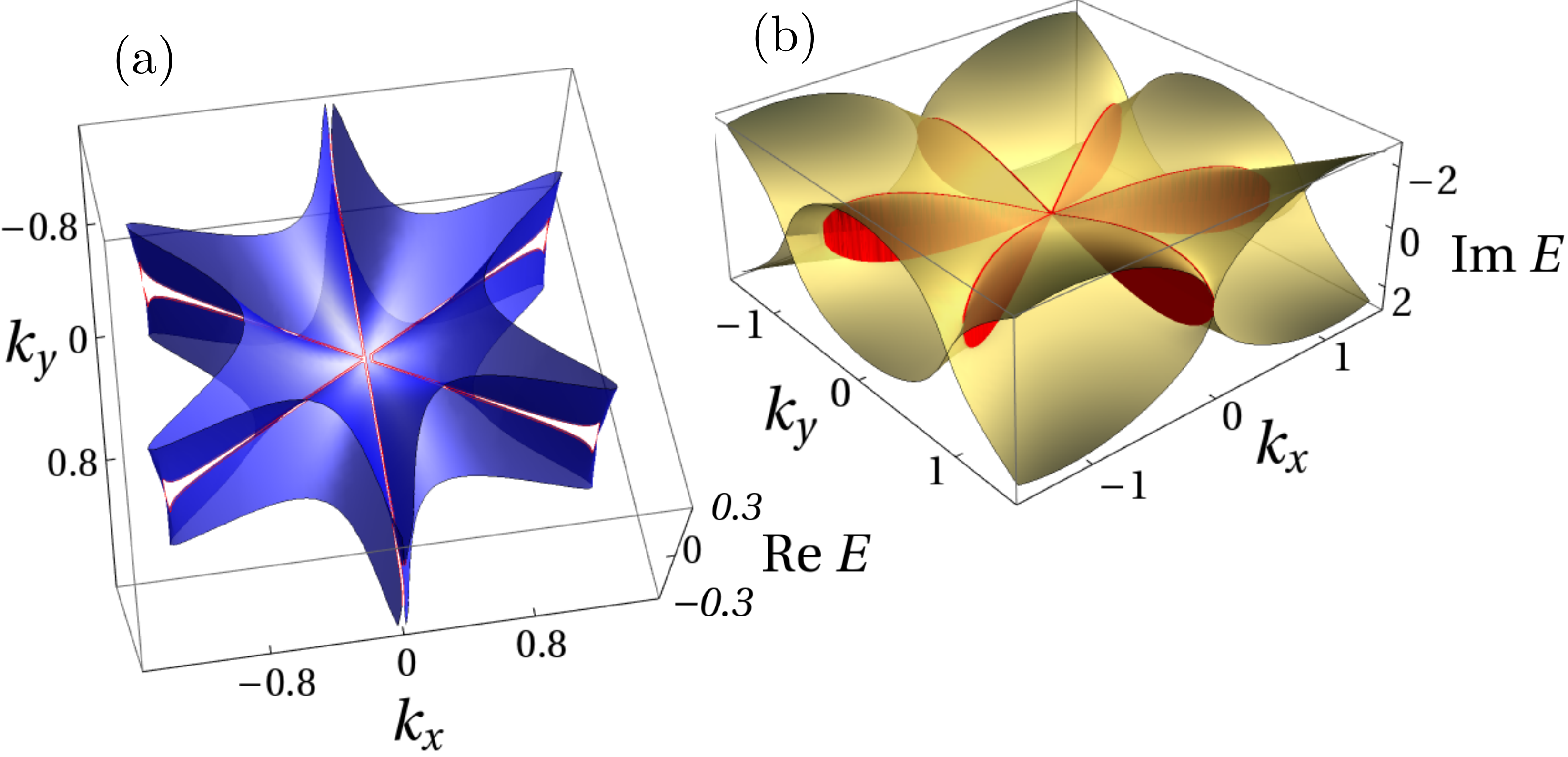}
	\caption{\textbf{Band diagrams showing HW in the presence of momentum dependent gain and loss.} (a) The real and (b) imaginary part of the energy spectrum are shown. We choose parameter values $\xi_x=1.5,$ $\xi_y=1.5,$ $\xi_z=0.0$, $k_z=0.0$, $v_z=1.0$ and $\eta=1.0$. The spectrum exhibits a hexagonal symmetry.} \label{fig1}	
\end{figure}
The NH effects on the topological aspects of the WSMs is a recent topic of interest \cite{Ghatak,Bergholtz,Gong,Luis,Luis1,Denner,zyuzin2018flat}. Numerous theoretical and experimental efforts \cite{f14,f15,f16,f17,f18,f19,f20,Banerjee,Our} have enhanced the understanding of different aspects of such systems in presence of NH perturbations. What makes the NH systems special is the existence of unique degenerate exceptional points (EPs), which are formed when both the eigenvalues as well as the eigenvectors coalesce \cite{Ghatak,Bergholtz,Cerjan1}. There may arise situations, where exceptional surfaces [exceptional contours (ECs)] appear in the system, containing many EPs.  It is to be noted that, in Weyl semimetals with charge $\pm 1$, the Weyl point converts into a Weyl exceptional ring in presence of a loss/gain along the $z$ direction. For Weyl semimetals having charge greater than $\pm 1,$ the exceptional contour need not appear in the form of ring \cite{Xu,yoshida2}, but it can take more complex shape satisfying the self-orthogonality condition. At the same time, quantum systems with light tuned properties are fascinating topic of current interest \cite{f1,f2,f3,f4,f5,f6,f7,f8,f9,f10,f11,f12,f13,f13a,f13b,f13c,f13d}. The light induced modifications of topological properties has led to the uncovering of various phenomena, which can cause important changes in the Fermi surface topology \cite{f13e,Our,banerjee2}. Recently, it has been shown that application of light can change the positions of the EPs in NH systems \cite{Our,banerjee2}. Besides, the light amplitude plays a crucial role in the merging or decoupling of ECs \cite{Our}. As such, driving is proposed to be a key controlling factor for tuning of Fermi surface topology and Berry charges of various topological systems.

\begin{figure*}	
\includegraphics[width=0.8\linewidth]{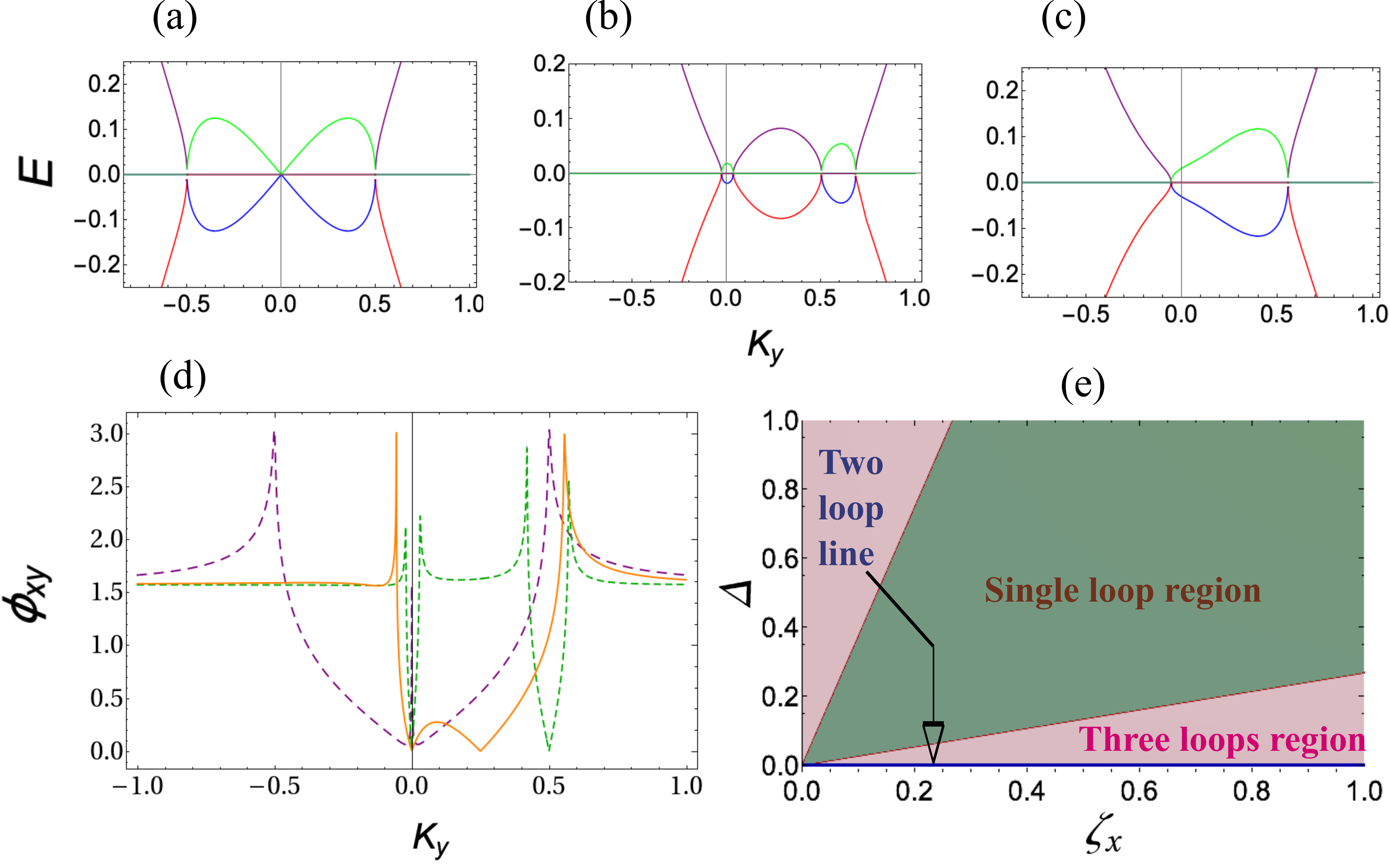}
\caption{\textbf{Band diagrams, winding number potential and phase diagram of the NH WSM under light illumination.} (a) Four EPs arising in the spectrum, where both the real and imaginary bands touch, in presence of a specific combination of gain/loss and without light ($A_0=0.0$ and $\zeta_x=0.5$) are shown. Two of the EPs are degenerate at the origin and the imaginary eigenbands form two loops as a consequence of complex $k^3$ coupling. (b) tunning both the light amplitude and warping parameter ($A_0=0.77$ and $\zeta_x=0.06$), the degenerate EPs split off and we get three asymmetric loops. (c) Beyond the critical value of the combination of $A_{0}$ and $\zeta_x$ ($A_0=0.5$ and $\zeta_x=0.25$) two of the generated EPs annihilate due to opposite topological winding (vorticity see Fig. 4) and spectrum shows a single loop. Real and imaginary part of the two bands of the spectrum are coloured with purple-red and blue-green, respectively. (d) The absolute value of the winding number potential $(\phi_{xy})$ for above three cases [purple ($A_0= \Delta=0.0$ and $\zeta_x=0.5$), green ($A_0=0.77$, $\Delta=A_0^{2}=0.5$ and $\zeta_x=0.06$) and orange ($A_0=0.5$, $\Delta=A_0^{2}=0.25$ and $\zeta_x=0.25$)] are shown. The divergences (sharp peaks) ensure the presence of EPs in the spectrum, signaling topological phase transitions. (e) The phase diagram showing different number of EPs [three (in a), four (in b) and two (in c)] as a function of $\Delta$ and $\zeta_x$. We have set $k_z=k_x=0.0$. }\label{fig2}
\end{figure*}

In this paper, starting from the DWSM, we show that properly choosing the loss/gain parameters can provide a hexagonal warping kind of effect in the system, whose origin is completely different. Unlike the original HW effect in Hermitian systems, the HW effect in non-Hermitian systems originate in the imaginary energy spectra. The importance of such a term is that in the presence of such HW the Fermi surface topology strikingly changes and it has a fundamental role on the nature and formation of the exceptional contours. The HW in this system has unique implications in the topology of the ECs. Hexagonally warped surface states are common in TIs and they contribute a real part in the eigen energies. In contrast, in case of a NH  DWSM, we show that not only is the origin of HW different, but also it adds an imaginary energy contribution to the band structure. This exceptional HW has a significant role in the formation, as well as merging of ECs. Further, it is a key ingredient to achieve tuning to different topological phases, whose topological invariants differ from the original phase, we start with. Apart from these unique features, the addition of driving field results in striking effects. We have presented a complete analysis of the topological invariants, i.e., the winding number and vorticity in presence of both the warping parameter and light amplitude. Finally, the charge distribution among the contours are discussed by analyzing the NH Berry curvature. Our results highlight the fascinating interplay between gain and loss induced HW and topology.

\begin{figure*}
	\includegraphics[width=0.8 \linewidth]{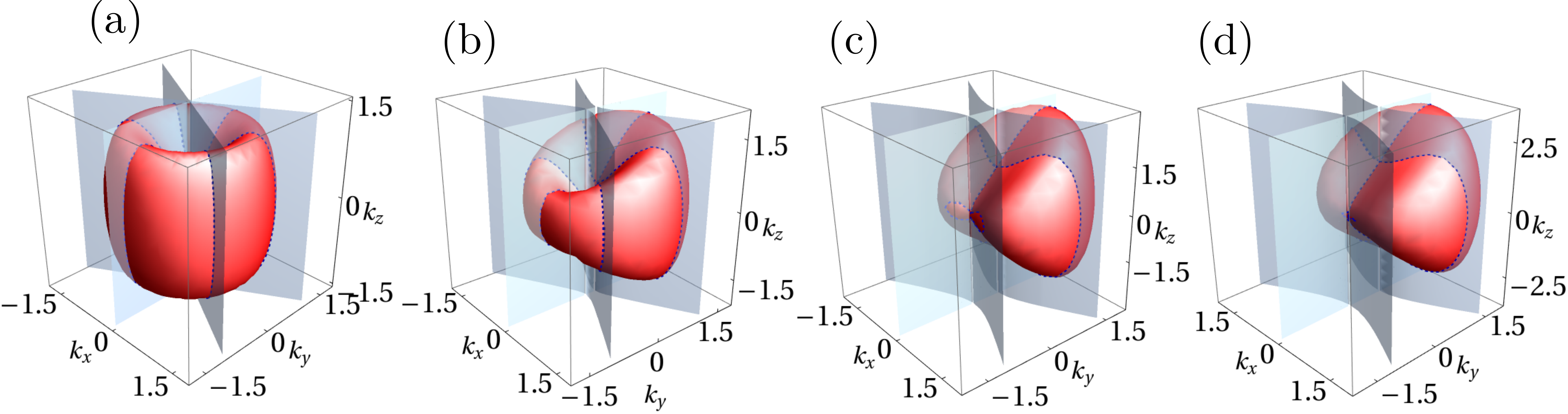}
\caption{\textbf{Light-tunable exceptional contours for NH WSMs.} The real (red) and imaginary (blue) exceptional surfaces are shown, and their intersection defines the exceptional contours (dotted blue line). (a) In the absence of light, tuning warping parameter ($A_0=0.0$ and $\zeta_x=1.5$), the contours show hexagonal symmetry and we obtain such six contours. By tuning both the parameters in (b) ($A_0=0.5$ and $\zeta_x=1.5$) these contours lose the hexagonal symmetry and tend to merge among them. (c) At critical values of light amplitude and warping parameter two of the contours merge ($A_0=0.7$ and $\zeta_x=1.5$), and (d) beyond this critical values of the combination ($A_0=0.98$ and $\zeta_x=1.5$) these two contours annihilate. We clearly see the evolution of contours by tuning the two important parameters accomplishing Lifshitz transitions. }\label{fig3}	
\end{figure*}
\section{The Hamiltonian and the energy bands in presence of driving}
We start with the Hamiltonian of a multi-WSM ($n$-WSM), which further includes a momentum dependent loss/gain term as follows
\begin{equation}
\label{Hamil}
H^{}(k)=\frac{1}{2m}\Big(k_{-}^{n}\sigma_{+}+k_{+}^{n}\sigma_{-}\Big)+\eta v_zk_z \sigma_{z}+i\bm{\zeta}(k)\cdot \bm{\sigma},
\end{equation}
where $k_{\pm} =k_{x}\pm i k_{y}$ and $\sigma_{\pm} =\sigma_{x}\pm i \sigma_{y}.$ The loss or gain vector $\bm{\zeta}(k)$ can be considered as
$\bm{\zeta}=\Big(\zeta_{x}k_{x},-\zeta_{x}k_{y},\zeta_{z}\Big).$ Here we consider the situation in which the amplitude of the $k_{x}$ direction gain $\zeta_x$ becomes equal to the loss amplitude in the $k_{y}$ direction (-$\zeta_x$). The energy eigen values (for $n=2$) in the $k_{z}=0$ plane are
\begin{align}\label{WOL1}
E^{0}_{\pm}&=\pm\Big[\Big(k^{2}-\zeta_{x}^{2}\Big) k_{}^2+i\zeta_{x}(k_{+}^{3}+k_{-}^{3})-\zeta_{z} ^2\Big]^{\frac{1}{2}},
\end{align}
where $k^{2}=( k_{x}^2+ k_{y}^2).$ It is important to note here that the second term in Eq. (\ref{WOL1}) reminds us of the HW in topological insulators \cite{Fu,Basak}. The origin of the $k^{3}$ coupling in Eq. (\ref{WOL1}) is solely due to the NH perturbation and provides an imaginary contribution to the eigen energies. In this sense, $\zeta_{x}$ may also be termed as the warping parameter. We have plotted the corresponding band structures in Fig.~\ref{fig1}. The imaginary $k^3$ spin orbit coupling mimicking term introduces a rotational hexagonal symmetry in the spectrum. One can visualize that due to the presence of this hexagonally symmetric term in the energy, an emergent symmetric band structure is obtained. Here $\zeta_z$ serves the role of a mass term and controls the gap between the imaginary bands.


Having understood the effect of momentum dependent gain and loss term, we next incorporate a periodic driving with a circularly polarized light and investigate it's role on the symmetries and ECs. The light is applied in the $x-z$ plane and thus $\hbar k_{i} \rightarrow \hbar k_{i}-eA_{i}$, where $\vec{A}(t+T)=\vec{A}(t)$, with $T=2\pi/ \omega$ as the periodicity. Here we consider circularly polarized light as a vector potential  $A(t)= A_{0}(\sin \omega t,0,\cos \omega t)$, where $A_{0}$ and $\omega$ are the amplitude and frequency of the driving optical field. It is to be noted here that introducing light changes the $k_x,k_z$ components, while $k_y$ remains unaffected. The full time-dependent Hamiltonian has the form
$H^{}(k,t) = H^{}(k) +H(t),$ where the time dependent part of the Hamiltonian is
\begin{align}\label{11}
H(t)&=\sin(\omega t)\Big[\frac{A_{0}}{m}\Big(k_{-}^{}\sigma^{+}+k_{+}^{}\sigma^{-}\Big)+i\zeta_{x}\sigma_{x}A_{0}\Big]\nonumber\\&+\eta v_{z} A_{0}\cos(\omega t)\sigma_{z}.
\end{align}
In the Floquet formalism, one can calculate the effective time independent Hamiltonian in the high frequency limit \cite{f1,f2,f6}, which yields
\begin{align}\label{12}
&H_{\mathrm{eff}}=\frac{1}{2m}\Big(\Big(k_{-}^{2}+i \Delta_{}k_{-}^{}\Big) \sigma_{+}+\Big(k_{+}^{2}-i \Delta_{}k_{+}^{}\Big)\sigma_{-}\Big)\nonumber\\&+\left(\eta v_z k_z+i\zeta_{z}\right)\sigma _z+i\zeta_{x}k_{x}\sigma_{x}+i\Big(\zeta_{x}k_{x}-\zeta_{x}\frac{\Delta}{2}\Big)\sigma_{y},
\end{align}
where $\Delta_{} =\frac{A_{0}^{2}v_{z}\eta}{\hbar \omega}.$
The corresponding energy eigen values for $k_{z}=0=\zeta_{z}$ are:
 \begin{align}\label{WOL2}
E^{}_{\pm}&=\pm\Big[\Big(k^{2}-\zeta_{x}^{2}\Big) k_{}^2+\frac{\Delta^2}{4}\Big(4k^{2}-\zeta_{x}^{2}\Big)-\Delta k_{y}\Big(\zeta_{x}^{2}+2k^2\Big)\nonumber\\&+i\zeta_{x}\Big(k_{+}^{3}+k_{-}^{3}+2\Delta k_{y}k_{x}+\Delta^{2}k_{x}\Big)-\zeta_{z} ^2\Big]^{\frac{1}{2}},
\end{align} 
which boils down to Eq. (\ref{WOL1}), when $\Delta=0.$ One important aspect here is that both the warping parameter $\zeta_{x}$ and the light amplitude $A_{0}$ are equally crucial for generating, tuning and merging of the contours. Thus, tuning either one (keeping the other fixed) or both the parameters can provide the desired exceptional physics in our system.

The energy dispersions are plotted with $k_{y}$ in Fig.~\ref{fig2} for a fixed value of $k_{x}=0.$  Four EPs arise in the spectrum where both the real and imaginary bands touch at a single point. In Fig.~\ref{fig2} (a), with the warping parameter $\zeta_{x} =0.5$ and $A_{0}=0$, we note that two EPs are degenerate at the origin and the imaginary eigenbands form two loops, which appear as a consequence of the complex $k^3$ coupling. In Fig.~\ref{fig2} (b), tuning both the parameters ($A_0=0.77,$ $\zeta_{x}=0.06$) the degenerate EPs split off, and we obtain three asymmetric loops. Beyond critical values of light amplitude ($A_0=0.5$) and $\zeta_{x}=0.25$, two of the EPs annihilate due to opposite topological winding (for vorticity, see discussion later) and the spectrum shows a single loop [Fig.~\ref{fig2} (c)]. Fig.~\ref{fig2} (d) presents the winding potential ($\phi_{xy}=\arctan\left(\frac{h_{x}}{h_{y}}\right),$ where the Hamiltonian can be written as $H_{eff}=h_{x}\sigma_{x}+h_{y}\sigma_{y}$) and will be discussed in due course. In Fig.~\ref{fig2} (e), we have shown the phase diagram of the number of EPs with light induced term $\Delta$ and warping parameter $\zeta_{x}.$ It can be seen that the combined action of $\Delta$ and $\zeta_{x}$ leads to different regions for obtaining different number of EPs. Here the green region shows the parameter space for which we have only two EPs, and the eigenvalues form only one loop and corresponds to Fig.~\ref{fig2} (c). Similarly, the two pink regions correspond to Fig.~\ref{fig2} (b). The blue line along the $\zeta_{x}$ axis shows exactly same number of EPs as is shown in Fig.~\ref{fig2}(a). This clearly highlights the role of both the parameters in the formation and control of EPs. The deciding factor about the number of Eps is the combined action of the key ingredients ($A_{0}$ and $\zeta_{x}.$) As is indicated in the plot \ref{fig2} (e), the region where the number of non-degenerate EPs are four is denoted by $(\zeta_{x}^{2}+\Delta^{2})> 4\zeta_{x}\Delta,$ which one finds by solving Eq. (\ref{WOL2}).  The region where $(\zeta_{x}^{2}+\Delta^{2})< 4\zeta_{x}\Delta$ is satisfied, contains two exceptional points. Else, for $\Delta=0,$ is the line where we have four Eps, among which two are degenerate at $k_{y}=0.$

\begin{figure}
\includegraphics[scale=0.28]{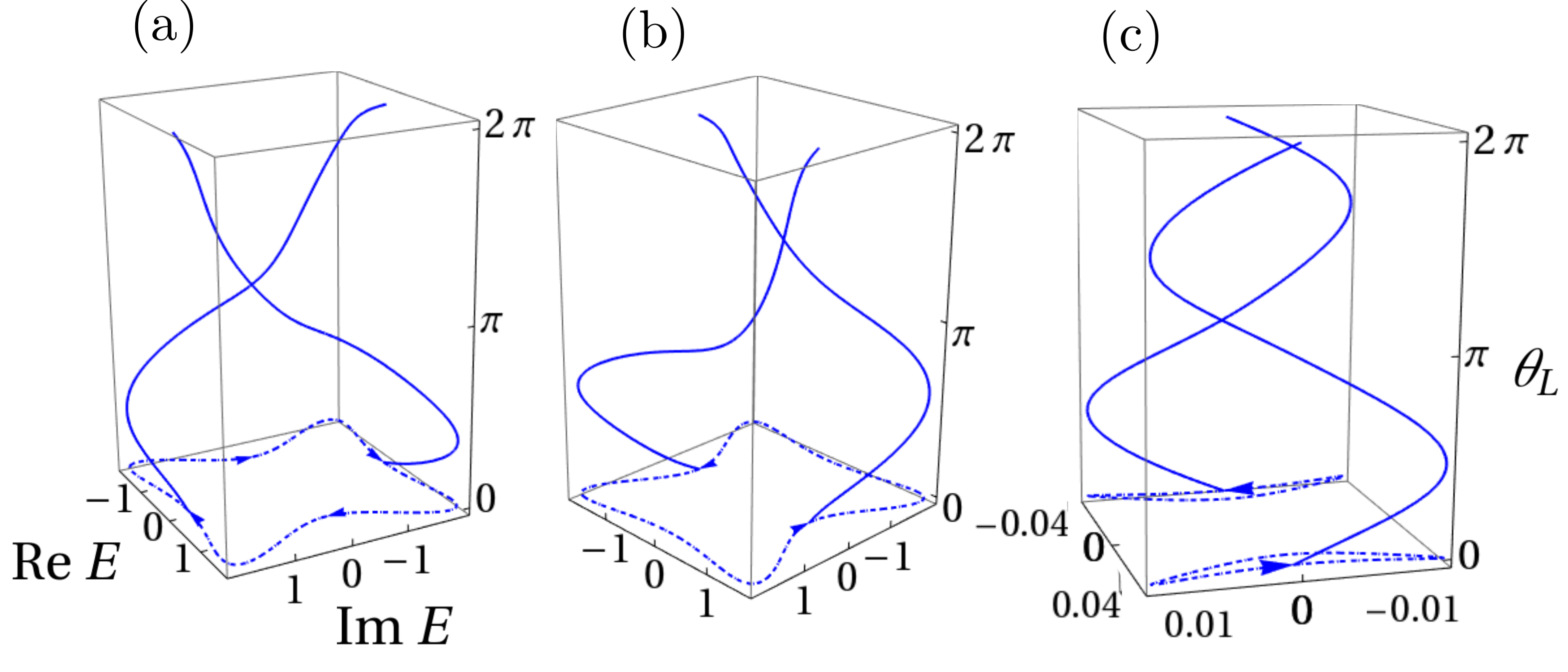}
\caption{\textbf{Illustration of the topological phase transitions from vorticity.} The evolution of two complex eigenbands when the contour parameterized by $\theta_L$ encircles (a) the left most and (b) right most EP in the lower panel of Fig.~\ref{fig5} (a) and (b). Their projections (dashed blue lines) onto the complex plane are shown. The orientations of two eigenbands are shown with the arrow. Note that the eigenbands wind around each other in opposite directions in the two cases. In (a) they wind clockwise, whereas in (b) they wind anti-clockwise. (c) The contour encircles the degenerate exceptional points at the origin in the lower panel of Fig.~\ref{fig5}. These two exceptional points have opposite winding and total vorticity vanishes. This clearly illustrates the swapping (or its lack thereof) of complex eigenbands and the resulting topological phase transitions.} 
\label{vorticity}
\end{figure} 

In Fig.~\ref{fig3}, the EC plots for the energy are presented for different values of $A_{0}$ and $\zeta_{x}$. In the absence of light ($A_{0}=0,$ $\zeta_{x}=1.5$) the contours show hexagonal symmetry and we achieve six symmetric contours [see Fig.~\ref{fig3} (a)]. Keeping $\zeta_{x}$ to be fixed and tuning the amplitude of light [Fig.~\ref{fig3} (b)] ($A_{0}=0.5$) the hexagonal symmetry starts distorting and the contours tend to merge among them. At a critical amplitude of light $A_{0}=0.7$, two of the contours merge [Fig.~\ref{fig3} (c)] and beyond this critical value, the two contours annihilate and we are left with four remaining contours [Fig.~\ref{fig3} (d)]. The evolution of contours with increasing light and for a fixed value of the warping parameter accomplishes the Lifshitz transition \cite{Our}.
\section{The Winding Number and Berry Curvature}
The topology of the hexagonally warped ECs can be well understood from an analysis of the winding number. If the Hamiltonian has a form, $H=h_{x}\sigma_{x}+h_{y}\sigma_{y},$  the winding number can be defined as~\cite{Gong,Yin}
\begin{align}\label{WN}
W_{}=\frac{1}{2\pi}\int_{-\infty}^{\infty} dk_{x} \partial{_{k_{x}}}\phi_{xy},
\end{align}
	where $\phi_{xy}=\arctan\left(\frac{h_{x}}{h_{y}}\right)$ is coined as the winding potential. In Ref. \cite{Our}, we have shown that a constant gain along $k_{z}$ leads to a winding of $+1/2$ for a double WSM. Here we analytically calculate the following values of the winding number
\begin{widetext}
\begin{align}\label{win}
W_{} &= -1,~~~\frac{(\Delta-\zeta_{x})}{2}-\sqrt{\frac{(\Delta-\zeta_{x})^{2}}{4}- \frac{\Delta \zeta_{x}}{2}}<|k_{y}|<\frac{(\Delta-\zeta_{x})}{2}+\sqrt{\frac{(\Delta-\zeta_{x})^{2}}{4}- \frac{\Delta \zeta_{x}}{2}} \nonumber\\
&=-\frac{1}{2},~~~\frac{(\Delta\pm\zeta_{x})}{2}\mp\sqrt{\frac{(\Delta\pm\zeta_{x})^{2}}{4}\pm \frac{\Delta \zeta_{x}}{2}}<|k_{y}|< \frac{(\Delta\mp\zeta_{x})}{2}\mp\sqrt{\frac{(\Delta\mp\zeta_{x})^{2}}{4}\mp \frac{\Delta \zeta_{x}}{2}}\nonumber\\
&=~~0,~~~ \mathrm{otherwise.}
\end{align}
\end{widetext}
\begin{figure*}
		\includegraphics[scale=0.3]{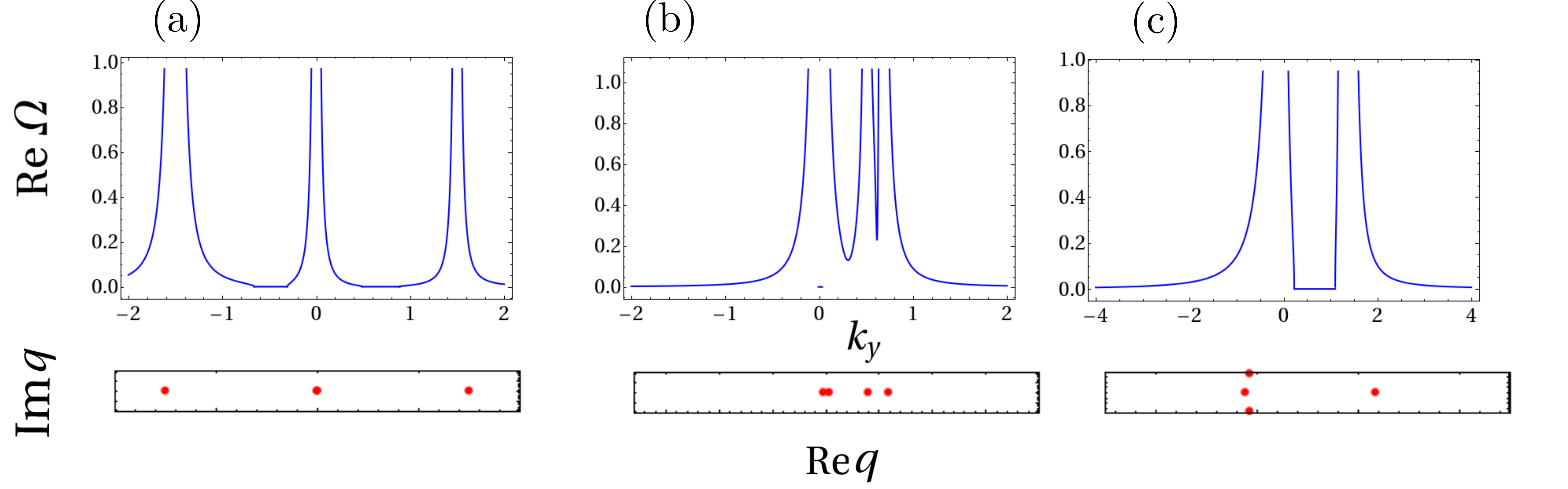}
		\caption{\textbf{Berry curvature density showing the topological charge distribution under light illumination.} (a) Normalized Berry curvature density as a function of the radial momentum without light in the presence of warping parameter ($\zeta_x=1.5$). The divergences signal the presence of NH band degeneracies (exceptional points) where the topological charge accumulates. In the absence of light we obtain three peaks, (b) while beyond the critical values of warping parameter and amplitude of light ($A_0=0.77$ and $\zeta_x=0.06$) the four peaks confirm the presence of four exceptional points in the spectrum. (c) Eventually, beyond these critical values two of the exceptional points annihilate for ($A_0=0.77$ and $\zeta_x=0.6$) and we achieve two peaks, which shows the topological charge distribution. The lower panel shows the location of EPs with light tuning. The imaginary co-ordinates of momentum signify their absence. } 
		\label{fig5}
\end{figure*} 
This demonstrates that the winding number can be readily controlled depending upon the values of the light induced term, $\Delta$ and the HW parameter, $\zeta_{x}.$ In Fig.~\ref{fig3} (d), we have plotted the winding potential, $\phi_{xy}$, with $k_{y}$ for three different values of the light amplitude and warping parameters and have observed that the interplay of both these parameters provides different number of peaks (four, three, two) in $\phi_{xy}.$ Importantly, the values we obtain for the winding number are interesting and need more explanation. The reason for obtaining the integer winding number for the non-Hermitian system under study is that, we compactify the line integral in Eq. (6) into a loop (since the derivative of $\phi_{xy}$ at -$\infty$ and $\infty$ are continuous) enclosing the exceptional points. Consequently, when we encircle a single exceptional point with the closed loop we obtain a half-quantized winding number. Whereas, if we enclose two exceptional points, the winding number turns out to be $\pm 1$ (with NH-Berry phase $\pi$), reminiscent of Hermitian topology. On the other hand, if no exceptional point is enclosed, winding number takes the trivial zero value.

Another important aspect, in connection to the complex energy spectrum, is that a new topological number, vorticity, can be defined, which effectively provides the information on the number of EPs, enclosed within a loop in a complex plane~\cite{Shen}. The complex eigenvalue for a single band is written as $E(\textbf k)=\mathopen|E(\textbf k)\mathclose|e^{i\theta_L(k)}$, where $\theta_L=\tan^{-1}({\mathrm{Im}E/\mathrm{Re}E})$. One of the unique features of NH systems is that it shows fractional vorticity~\cite{Shen}. The vorticity for the double WSM is presented in Fig.~\ref{vorticity}(a)-(c). If an odd number of EPs are within the selected closed contour, band swapping is visualized. In contrast, an integer or zero value would result if an even number of EPs are enclosed~\cite{banerjee2}. Our analysis confirms this picture.

Next, we analyze the NH Berry curvature [${\bf\Omega}_{LR}({\textbf k})= \nabla \times  {\cal B}^{LR}({\textbf k})$], in presence of both the warping and the driving parameters. Here ${\cal B}^{LR}({\textbf k})$ is the Berry gauge field, which is obtained from the left and right eigenvectors ($\psi^{L/R}$) of the Hamiltonian as ${\cal B}^{LR}(k)=i\left<\psi^{L}(k)|\nabla|\psi^{R}(k)\right>$\cite{Our}. The well-known relation between the Berry charge and curvature is ${\mathfrak C}=\int_{C} {\bf \Omega}_{LR}({\textbf k})\cdot d{\bf S}.$ Thus, knowing the nature of the Berry curvature one can comment on the topological charge distribution of the system. Fig.~\ref{fig5} shows the plot of the real part of the Berry curvature with the radial momentum $k_{\rho}.$ It is to be noted here that we consider the cylindrical coordinate with $k_{x}=k_{\rho}\cos(\phi)$ and $k_{y}=k_{\rho}\sin(\phi).$ Next we consider $\phi=\pi/2,$ which provides $k_{x}=0$ and $k_{y}=k_{\rho}.$ In Fig.~\ref{fig5} (a) the Berry curvature density diverges at three different positions for $A_{0}=0$ and warping parameter $\zeta_{x}=1.5.$ The diverging Berry curvature is associated to the exceptional degeneracy. Importantly, at $k_{\rho}(k_{y})=0,$ two EPs are present together, in the absence of light. Tuning both the parameters ($A_{0}$ and $\zeta_{x}$) judiciously yields four non-degenerate EPs [see Fig.~\ref{fig5} (b)] for $A_{0}=0.77$ and $\zeta=0.06$. In Fig.~\ref{fig5} (c), further modification of the two parameters ($A_0=0.77$ and $\zeta_x=0.6$) compels two of the EPs to annihilate and we are left with only two peaks. The lower panel of Fig.~\ref{fig5} shows the locations of EPs with different light amplitudes and is consistent with our Berry curvature analysis.
\section{Discussion and CONCLUSION} 
Finally, we would like to shed some light on the experimental feasibility of our results. In recent years, there have been several ingenious experiments suggested as well as realized to investigate the exceptional physics by tuning the gain/loss terms judiciously, in diverse physical settings, including photonic \cite{Chen} and acoustic metamaterials\cite{yosida1,Ghatak1}, cold atomic gases \cite{xu2017weyl}, heavy-fermion systems \cite{yoshida} and even in a TI-ferromagnet junctions \cite{Emil}. Complex momentum dependent coupling can be realized in a spectral photonic lattice where lattice sites are represented by discrete frequency channels driven by non-linear interaction from stronger pump lasers \cite{Bryn}. In this setting, controllable complex coupling can be obtained by tuning the spectrum of the optical pump. An alternative approach to the possible experimental realization of our proposal is by implementing imaginary gauge field in coupled two resonators with an engineered anti-resonance ring, thus allowing for directional coupling \cite{Midya}. Another possible direction to experimentally realize our results is to consider a topo-electrical circuits where NH coupling can be realized by resistively connecting different nodes in the circuit \cite{Our,Xio}. The experimental detection of nodal band structures is possible by tracing the complex admittance spectra, which shows striking changes at the EPs and thus signals their presence \cite{Tobi}. Interestingly, very recently Weyl exceptional rings were realized in an evanescently coupled bipartite optical waveguide array by introducing tunable breaks in the waveguide, which lead to loss in the system \cite{Cerjan}. Unique Fermi arcs and topological charge distributions were also demonstrated in a controllable manner in this set up. In Ref. \cite{Woodpole,Cerjan1}, authors have proposed using metallic chiral woodpile photonic crystal consisting of layers of a hexagonal lattice designed to operate in the terahertz frequency band and allowing the on-site complex energies in each lattice layer, to observe ECs. As the NH terms can be controlled in various readily available topological systems, we are optimistic that our proposals can be directly verified in experiments in the near future.

In summary, we have studied the role of momentum dependent loss/gain on the Fermi surface topology of DWSM, which yields completely new phenomenology in the arena of the DWSMs. We showed that a combination of the two tuning parameters, light amplitude and warping strength, can provide striking novel features related to ECs. One can visualize generation of non-degenerate ECs and merging of the same. We have also discussed physics related to winding number of the system. The topological charge distribution of the generated contours was also analyzed, which demonstrated the possibility of controlling the EPs by tuning these parameters.

\section{Acknowledgments} D.C. acknowledges financial support from DST (project number SR/WOS-A/PM-52/2019). A.B. is supported by the Prime Minister's Research Fellowship (PMRF). A.N. acknowledges support from a startup grant (SG/MHRD-19-0001) of the Indian Institute of Science and DST-SERB (project number SRG/2020/000153).


\begin{thebibliography}{999}
		\bibitem{Hasan} M. Z. Hasan and C. L. Kane, Rev. Mod. Phys.
		{\bf 82}, 3045 (2010).
		\bibitem{Fu} L. Fu, Phys. Rev. Lett. {\bf 103}, 266801 (2009).
		\bibitem{Basak} S. Basak, H. Lin, L. A. Wray, S.-Y. Xu, L. Fu, M. Z. Hasan, and A. Bansil, Phys. Rev. B 84, 121401(R), (2011).
		\bibitem{Vafek} O. Vafek and A. Vishwanath, Annu. Rev. Condens. Matter Phys. {\bf 5}, 83 (2014).
		\bibitem{Kane} C. L. Kane and E. J. Mele, Phys. Rev. Lett.  {\bf  95}, 226801 (2005).
		\bibitem{Burkov} A. A. Burkov, Nature Materials {\bf 15}, 1145 (2016).
		\bibitem{Armitage} N. P. Armitage, E. J. Mele, and A. Vishwanath, Rev.
		Mod. Phys. {\bf 90}, 015001 (2018).
		\bibitem{Gupta} Amit Gupta, arXiv:1703.07271.
		\bibitem{MW1} C. Fang, M. J. Gilbert, X. Dai, and B. Andrei Bernevig, PRL {\bf 108}, 266802 (2012).
		\bibitem{TW} Shi-Xin Zhang, Shao-Kai Jian, and Hong Yao, Phys. Rev. B {\bf 96}, 241111(R) (2017).
		\bibitem{Fang}Chen Fang, Matthew J. Gilbert, Xi Dai, and B. Andrei Bernevig, Phys. Rev. Lett. {\bf 108}, 266802 (2012).
		\bibitem{Xu1} Gang Xu, Hongming Weng, Zhijun Wang, Xi Dai, and Zhong Fang, Phys. Rev. Lett. {\bf 107}, 186806 (2011).
		\bibitem{Huang} S.-M. Huang, Su-Y. Xu , I. Belopolski, Chi-C. Lee, G. Chang, Tay-Rong Chang, B. Wang, N. Alidoust, G. Bian, M. Neupane, D. Sanchez, H. Zheng , H.-T. Jeng, A. Bansil, T. Neupert, H. Lin, M Zahid Hasan, PNAS, {\bf 113} (5), 1180 (2016).
		\bibitem{Zhang}S.-X. Zhang, S.-K. Jian, and H. Yao, Phys. Rev. B {\bf 96}, 241111(R) (2017).
	\bibitem{Cerjan1} A. Cerjan, M. Xiao, L. Yuan, and S. Fan, Phys. Rev. B {\bf 97}, 075128 (2018).	
		\bibitem{Ghatak} A. Ghatak and T. Das, J. Phys.: Condens. Matter {\bf 31}, 263001 (2019).
		\bibitem{Bergholtz} E. J. Bergholtz, J. C. Budich, and F. K. Kunst,	Rev. Mod. Phys. {\bf 93}, 15005 (2021).
		\bibitem{Gong} Z. Gong, Y. Ashida, K. Kawabata, K. Takasan, S. Higashikawa,
		and M. Ueda, Phys. Rev. X {\bf 8}, 031079 (2018).
		\bibitem{Luis} Luis E F Foa Torres, Journal of Physics: Materials, {\bf 3}, 1 (2019).
		\bibitem{Luis1} V. M. Alvarez, J. B. Vargas, M. Berdakin, and L. F. Torres,
		Eur. Phys. J.: Spec. Top. {\bf 227}, 1295 (2018).
		\bibitem{f14}L. Zhou, J. Gong, Physical Review B {\bf 98}, 205417,(2018). 
		\bibitem{f15}L Zhou, J Pan, Physical Review A  {\bf 100}, 053608, (2019).
		\bibitem{f16}L Zhou, Physical Review B  {\bf 100}, 184314,  (2019).
		\bibitem{f17}L Zhou, Physical Review B  {\bf 101}, 014306, (2020).
		\bibitem{f18}J Pan, L Zhou, Physical Review B {\bf 102}, 094305, (2020).
		\bibitem{f19}L Zhou, Y Gu, J Gong, arXiv:2009.13078.
		\bibitem{f20} S. Jana, D. Chowdhury, A. Saha, arXiv:2010.02852.	
	\bibitem{Banerjee} A. Banerjee and A. Narayan, J. Phys.: Condens. Matter {\bf 33}, 225401, (2021).
		\bibitem{Our} D. Chowdhury, A. Banerjee, and A. Narayan, Phys. Rev. A {\bf 103}, L051101, (2021).
			\bibitem{f1} T. Oka and H. Aoki, Phys. Rev. B {\bf 79}, 081406(R) (2009).
		\bibitem{f2} J. Cayssol, B. Dóra, F. Simon, and R. Moessner, Phys. Status
		Solidi RRL {\bf 7}, 101 (2013).
		\bibitem{f3} N. H. Lindner, G. Refael, and V. Galitski, Nat. Phys. {\bf 7}, 490 (2011).
		\bibitem{f4} M. C. Rechtsman, J. M. Zeuner, Y. Plotnik, Y. Lumer, D.
		Podolsky, F. Dreisow, S. Nolte, M. Segev, and A. Szameit, Nature (London) {\bf 496}, 196 (2013). 
		\bibitem{f5} Y. Wang, H. Steinberg, P. Jarillo-Herrero, and N. Gedik, Science, {\bf 342}, 453 (2013).
		\bibitem{f6} M. S. Rudner and N. H. Lindner, arXiv:2003.08252.
		\bibitem{f7} A. Narayan, Phys. Rev. B {\bf 91}, 205445 (2015).
		\bibitem{f8} Z. Yan and Z. Wang, Phys. Rev. Lett. {\bf 117}, 087402 (2016).
		\bibitem{f9} Ching-Kit Chan, Yun-Tak Oh, Jung Hoon Han, and Patrick A. Lee, Phys. Rev. B {\bf 94}, 121106 (2016).
		\bibitem{f10} A. Narayan, Phys. Rev. B {\bf 94}, 041409(R) (2016).
		\bibitem{f11} K. Taguchi, D. Xu, A. Yamakage, and K. T. Law, Phys. Rev. B {\bf 94}, 155206 (2016).
		\bibitem{f12} Z. Yan and Z. Wang, Phys. Rev. B {\bf 96}, 041206(R) (2017).
		\bibitem{f13} R. Jaiswal and A. Narayan, Phys. Rev. B {\bf 102}, 245416 (2020).
			\bibitem{f13a}J. W. McIver, B. Schulte, F.-U. Stein, T. Matsuyama, G. Jotzu, G. Meier and A. Cavalleri 
		Nature Physics  {\bf 16}, 38(2020).
		\bibitem{f13b}S. A. Sato, J. W. McIver, M. Nuske, P. Tang, G. Jotzu, B. Schulte, H. Hübener, U. De Giovannini, L. Mathey, M. A. Sentef, A. Cavalleri, and A. Rubio, Phys. Rev. B  {\bf 99}, 214302(2019).	
		\bibitem{f13c}Hannes Hübener, Michael A. Sentef, Umberto De Giovannini, Alexander F. Kemper and Angel Rubio 
		Nature Communications, {\bf 8}, 13940 (2017). 
		\bibitem{f13d} Edbert J. Sie, Clara M. Nyby, C. D. Pemmaraju, Su Ji Park, Xiaozhe Shen, Jie Yang, Matthias C. Hoffmann, B. K. Ofori-Okai, Renkai Li, Alexander H. Reid, Stephen Weathersby, Ehren Mannebach, Nathan Finney, Daniel Rhodes, Daniel Chenet, Abhinandan Antony, Luis Balicas, James Hone, Thomas P. Devereaux, Tony F. Heinz, Xijie Wang and Aaron M. Lindenberg. Nature {\bf 565}, 61 (2019).
		\bibitem{f13e}Robert J. Kirby, Lukas Muechler, Sebastian Klemenz, Caroline Weinberg, Austin Ferrenti, Mohamed Oudah, Daniele Fausti, Gregory D. Scholes, and Leslie M. Schoop
		Phys. Rev. B \textbf{103}, 205138, (2021).
		\bibitem{banerjee2} A. Banerjee and A. Narayan, Phys. Rev. B {\bf 102}, 205423 (2020).
			\bibitem{Yin} C. Yin, H. Jiang, L. Li, R. Lü, and S. Chen,
		Phys. Rev. A {\bf 97}, 052115, (2018).
		\bibitem{Shen} H. Shen, Bo Zhen, and L. Fu, Phys. Rev. Letts {\bf 120}, 146402 (2018).
		\bibitem{Chen} Weijian Chen, Şahin Kaya Özdemir, Guangming Zhao, Jan Wiersig and Lan Yang 
Nature {\bf 548}, 192,(2017).
	\bibitem{yosida1} T. Yoshida and Y. Hatsugai, Phys. Rev. B {\bf 100}, 054109 (2019).
	\bibitem{Ghatak1} A. Ghatak, M. Brandenbourger, J. van Wezel, C. Coulais, Proc. Natl. Ac. Sc. U.S.A. \textbf{117} (47) 29561-29568 (2020).
\bibitem{Xu} Yong Xu, Sheng-Tao Wang, and L.-M. Duan, Phys. Rev. Letts, {\bf 118}, 045701 (2017).
\bibitem{yoshida2} T. Yoshida, R. Peters, N. Kawakami, and Y. Hatsugai, Phys. Rev. B \textbf{99}, 121101(R), (2019).
\bibitem{Emil} E. J. Bergholtz and J.C. Budich, Phys. Rev. Research {\bf 1}, 012003(R), (2019).
\bibitem{Bryn} B. A. Bell, K. Wang, A. S. Solntsev, D. N. Neshev, A. A. Sukhorukov, and B. J. Eggleton, {\bf 4}, 1433, (2017).
\bibitem{Midya} B. Midya, H. Zhao, and L. Feng, Nat. Commun. {\bf 9}, 1 (2018).
\bibitem{Xio} X-X Zhang and M. Franz, Phys. Rev. Lett. {\bf 124}, 046401, (2020).
\bibitem{Tobi} T. Hofmann, T. Helbig, F. Schindler, N. Salgo, M. Brzezińska, M. Greiter, T. Kiessling, D. Wolf, A. Vollhardt, A. Kabaši, C. Hua Lee, A. Bilusic, R. Thomale, and T. Neupert, Phys. Rev. Research 2, 023265
\bibitem{Cerjan} A. Cerjan, S. Huang, M. Wang, K. P. Chen, Y. Chong  and M. C. Rechtsman, Nature Photonics,{\bf 13}, 623 (2019).
\bibitem{Woodpole} Ming-Li Chang, Meng Xiao, Wen-Jie Chen, and C. T. Chan, Phys. Rev. B \textbf{95}, 125136 (2017).

\bibitem{Denner} M. Michael Denner, Anastasiia Skurativska, Frank Schindler, Mark H. Fischer, Ronny Thomale, Tomáš Bzdušek and Titus Neupert, Nat. Commun. {\bf 12}, 5681 (2021).
\bibitem{zyuzin2018flat} Zyuzin, Alexander A and Zyuzin, A Yu, Phys. Rev. B \textbf{97}, 041203(R) (2018).
\bibitem{yoshida} T. Yoshida, R. Peters, and N. Kawakami, Phys. Rev. B \textbf{98}, 035141 (2018).
\bibitem{xu2017weyl} Xu, Yong and Wang, Sheng-Tao and Duan, L-M, Phys. Rev. Lett. {\bf 118}, 045701, (2017).
\bibitem{yoshida2} T. Yoshida, R. Peters, N. Kawakami, and Y. Hatsugai, Phys. Rev. B \textbf{99}, 121101(R), (2019).
	\end{thebibliography}
\end{document}